\documentclass[12pt]{article}
\usepackage{epsfig}

\def\beq{\begin{equation}}
\def\eeq#1{\label{#1}\end{equation}}
\def\eeqn{\end{equation}}

\def\beqa{\begin{eqnarray}}
\def\eeqa#1{\label{#1}\end{eqnarray}}
\def\eeqan{\end{eqnarray}}

\let\bar=\overbar

\def\Dslash{\not{\hbox{\kern-4pt $D$}}}
\def\dslash{\not{\hbox{\kern-2pt $\del$}}}

\def\msb{{\bar{\ssstyle M \kern -1pt S}}}

\def\lQ{\Lambda_{\rm QCD}}
\def\siml{{\ \lower-1.2pt\vbox{\hbox{\rlap{$<$}\lower6pt\vbox{\hbox{$\sim$}}}}\ }}
\def\simg{{\ \lower-1.2pt\vbox{\hbox{\rlap{$>$}\lower6pt\vbox{\hbox{$\sim$}}}}\ }}
\newcommand{\MS}{\overline{\rm MS}}

\def\als{\alpha_{\rm s}}
\def\siml{{\ \lower-1.2pt\vbox{\hbox{\rlap{$<$}\lower6pt\vbox{\hbox{$\sim$}}}}\ }}

\def\Title#1{\begin{center} {\Large {\bf #1} } \end{center}}

\begin{document}

\Title{NRQCD and Quarkonia}

\begin{center}{\large \bf Contribution to the proceedings of HQL06,\\
Munich, October 16th-20th 2006}\end{center}

\bigskip\bigskip


\begin{raggedright}  

{\it Nora Brambilla\index{Brambilla, N.}\\
Dipartimento di Fisica \\
Universit\`a di Milano and  INFN,\\ 
via Celoria 16,\\ 
20133 Milan, ITALY}
\bigskip\bigskip
\end{raggedright}

\section{Introduction}
Quarkonia play an important role in several 
high energy experiments. The diversity, quantity 
and accuracy of the data still under analysis and currently being collected is impressive 
and includes:  data on quarkonium formation from BES at the Beijing Electron Positron Collider
 (BEPC), E835 at Fermilab, and CLEO at the Cornell  Electron Storage Ring (CESR); 
clean samples of charmonia produced in B-decays, in photon-photon
fusion and in initial state radiation, at the B-meson factories, 
BaBar at PEP-II and Belle at KEKB, including the unexpected
observation of large amounts  of associated $(c\overline{c})(c\overline{c})$ production
 and the observation of new and possibly exotics quarkonia states;
the CDF and D0 experiments at Fermilab measuring heavy quarkonia
production from gluon-gluon fusion in $p\bar{p}$ annihilations at
2~TeV; the Selex experiment at Fermilab with  the preliminary observation of 
possible candidates for doubly charmed baryons; 
ZEUS and H1, at DESY, studying charmonia production in photon-gluon
fusion; PHENIX and STAR, at RHIC, and NA60, at CERN, studying charmonia
production, and suppression, in heavy-ion collisions.
This has led to the discovery of new states, new production mechanisms, new decays and transitions,
and in general to the collection of high statistics and precision data sample.
In the near future, even larger data samples are expected from the
BES-III upgraded experiment, while the B factories and the
Fermilab Tevatron  will continue to supply
valuable data for few years.  Later on, new experiments at new facilities will become
operational (the LHC experiments at CERN, Panda at GSI, hopefully a Super-B factory,
a Linear Collider, etc.) offering fantastic challenges and opportunities in this field.  
A  comprehensive review of the experimental and theoretical 
status of heavy quarkonium physics may be found in the Cern Yellow Report prepared by 
the Quarkonium Working Group 
~\cite{Brambilla:2004wf}. Many excellent reviews of the field have been presented at this meeting
~\cite{here}.

On the theory side, systems made by two heavy quarks are a rather unique laboratory.
They are characterized by the existence of a hierarchy of energy scales in correspondence 
of which one can construct a hierarchy of nonrelativistic effective field theries (NR EFT),
each one with  less degrees of freedom left dynamical and thus simpler. Some of these 
physical scales are large and  may be treated in perturbation theory. 
The occurrence of these two facts makes two heavy quark systems accessible in QCD. 
In particular, the factorization of high and low energy scales realized 
in the EFTs allows us to study low energy QCD effects in a systematic way.
Today the remarkable progress in the construction of these nonrelativistic EFTs
together with the advance in lattice QCD give us  well based theory tools to investigate 
heavy quarkonia.

Therefore, on the one hand the progress in our understanding of EFTs makes it 
possible to move beyond phenomenological models and to  provide a systematic description 
from QCD of several aspects of heavy-quarkonium physics. On the other hand, the recent 
progress in the measurement of several heavy-quarkonium observables makes it meaningful 
to address the problem of their precise theoretical determination.
In this situation heavy quarkonium becomes a very special and relevant system to advance 
our understanding of strong interactions and our control of some parameters of the Standard
Model. 

Here I will briefly review some of the recent developments in 
the construction of NR EFTs with the main emphasis 
on the physical applications. For some  reviews 
see ~\cite{Brambilla:2004jw,nreftrev,nreftrev2,nreftrev3}.

\section{Scales and Effective Field Theories}
The description of hadrons containing 
two heavy quarks is a rather challenging problem,  
which adds to the complications of the bound state in field theory 
those coming from the nonperturbative QCD low-energy dynamics.
A  simplification is provided  by  the nonrelativistic nature 
suggested by the large mass of the heavy quarks
and  manifest in the spectrum pattern.
As nonrelativistic systems, quarkonia are  characterized
 by three energy scales, hierarchically ordered by the heavy quark velocity
in the center of mass frame 
$v \ll 1$: the mass $m$ (hard scale), the momentum transfer $mv$ (soft scale), 
which is proportional to the inverse of the typical size of the system $r$,
 and  the binding energy $mv^2$ (ultrasoft scale), which is proportional to the 
inverse of the typical 
time of the system. In bottomonium $v^2 \sim 0.1$, in charmonium $v^2 \sim 0.3$,
 in $t\bar{t}$  $v\sim 0.15$.
In perturbation theory $v\sim \als$. Feynman diagrams will get contributions 
from all momentum regions associated with these scales.
Since these momentum regions depend on $\als$, each Feynman diagram contributes 
to a given observable with a series in $\als$ and a non trivial counting.
Besides, the  $\als$ associated to different momentum region are evaluated at 
different scales.
For energy scales close to $\lQ$, the scale at which 
nonperturbative effects become dominant, perturbation theory breaks down 
and one has to rely on nonperturbative methods.
 Regardless of this, the non-relativistic hierarchy 
$ m \gg mv \gg mv^2,\,  m\gg \lQ $ will persist  also below  the  $\lQ$  threshold.

The wide span of involved  energy scales  makes also a lattice calculation in full QCD 
extremely challenging.  However, it is possible to  exploit  the existence of a hierarchy of scales  
by introducing a hierarchy of nonrelativistic effective field theories.
Lower energy EFTs may be constructed by systematically integrating out 
modes associated to  energy scales not relevant for the two quark  system.
Such integration  is made  in a matching procedure that 
enforces the equivalence between QCD and the EFT at any  given 
order of the expansion in $v$. Any prediction of the EFT is therefore a 
prediction of QCD with an error of the size of the neglected order in $v$.
By integrating out the hard modes, one  obtains Nonrelativistic QCD
 \cite{Caswell:1985ui,Bodwin:1994jh,Manohar:1997qy}.
 In such EFT, soft 
and ultrasoft scales are left dynamical and still their mixing  complicates 
 calculations and power counting. 
In the last few years the problem of systematically treating the remaining
dynamical scales in an EFT framework has been addressed  by several groups
\cite{group} and has now reached a good level of understanding.
So one can go down one step further and integrate out also the soft scale 
in a matching procedure to the lowest energy and simplest EFT that can be 
introduced for quarkonia,
where only ultrasoft degrees of freedom remain dynamical.
Here I will review  potential NRQCD~\cite{Pineda:1997bj,Brambilla:1999xf},
for an alternative and equivalent EFT (in the case in which $\lQ$ is 
the smallest scale) see \cite{vnrqcd}.
In the case in which the soft scale is of the same order of $\lQ$, the matching 
to pNRQCD is still possible but it is nonperturbative.

\section{NonRelativistic QCD (NRQCD)}

NRQCD \cite{Caswell:1985ui,Bodwin:1994jh}
 is the EFT for two heavy quarks that follows from QCD by integrating out the hard 
scale $m$. Only the upper (lower) components  of the Dirac fields remain relevant
for dymanical quarks (antiquarks) at energies lower than $m$. Thus quark and antiquarks are 
described in terms of two-components Pauli spinor fields. 
The part of the NRQCD Lagrangian bilinear  in the heavy quark fields  is the same as 
Heavy Quark Effective Field Theory (HQET) (for a review see \cite{Neubert:1993mb})
but for the case of two heavy quarks also four fermion operators have to be considered.
The Lagrangian is organized as an expansion in $v$  and 
  $\als(m)$:
\begin{equation}
 {\cal L}_{\rm NRQCD}  = \sum_n  c_n(m,\mu)  \times  O_n(\mu, mv,mv^2,  \lQ)/m^n . 
\label{eftnrqcd}
\end{equation}
The NRQCD matching  coefficients 
$c_n$ are series in $\als $ and encode the ultraviolet physics that  
has been integrated out from QCD. The low energy operators $O_n$  are
constructed out of two or four  
  heavy quark/antiquark fields plus gluons. They are  counted 
in powers of $v$. Since two scales, soft and the ultrasoft, are  dynamical,
the power counting in $v$  is not unambiguous.
The imaginary part of the coefficients of the 4-fermion operators
contains the information on heavy quarkonium annihilations.
The NRQCD heavy quarkonium Fock state is 
given by a series of terms, increasingly subleading, where the leading term is a  $Q\bar{Q}$ 
in a color singlet state and the first correction, suppressed in $v$,
comes from a $Q\bar{Q}$ in a color  octet state plus a gluon.
The NRQCD Lagrangian can be used for studies of spectroscopy (on the lattice),
inclusive decays and electromagnetics threshold  production of heavy quarkonia.

\section{potential NonRelativistic QCD (pNRQCD)}
pNRQCD \cite{Pineda:1997bj,Brambilla:1999xf,Brambilla:2004jw}
 is the EFT for two heavy quark systems that follows from NRQCD 
by integrating out the soft scale $mv$. Here the role of the potentials and the
 quantum mechanical nature of the problem are realized in  the fact that the Schr\"odinger 
equation appears as zero order problem for the two quark states.
We may distinguish two situations:
1) weakly coupled pNRQCD when $mv  \gg \lQ$, 
where the matching from NRQCD to pNRQCD may be performed in
 perturbation theory;
2) strongly coupled pNRQCD when $ mv \sim \lQ$, 
where the matching has to be nonperturbative.
Recalling that $r^{-1}
  \sim mv$, these two situations correspond  to systems with inverse
typical radius smaller than or  of the same order as  $\Lambda_{\rm QCD}$.

\subsection{Weakly coupled pNRQCD}
The effective degrees of
freedom that remain dynamical  are: 
low energy $Q\bar{Q}$ (Pauli spinor) states that can be decomposed into  a singlet
 field $S$ and an octet field $O$  under colour transformations,
have  energy of order  $ \Lambda_{\rm QCD},\, mv^2$ and
    momentum   ${\bf p}$ of order $mv$; 
   low energy (ultrasoft (US))  gluons $A_\mu({\bf R},t)$ 
with energy  and momentum of order
$\lQ,\, mv^2$. 
All the  gluon fields are multipole  expanded (i.e.  expanded  in
the quark-antiquark distance $r$, $R$ being the center of mass). 
The Lagrangian is then given by terms of the type
\begin{equation}
{c_k(m, \mu) \over m^k}  \times  V_n(r
\mu^\prime, r\mu)   \times  O_n(\mu^\prime, mv^2, \lQ)\; r^n  .
\end{equation}
where the matching coefficients $c_k$ are inherited from NRQCD and 
contain the logs in the quark masses, the pNRQCD  
potential matching  coefficients $V_n$  encode the non-analytic behaviour in $r$
and the low energy operators $O_n$ are constructed in terms of singlet, octet fields and 
ultrasoft gluons.
At  leading order in the multipole expansion,
the singlet sector of the Lagrangian gives rise to equations of motion of the 
Schr\"odinger type. 
Each term in  the pNRQCD Lagrangian has a definite power counting.
The bulk of the interaction is carried by potential-like terms, but non-potential
interactions, associated with the propagation of low energy degrees of freedom
are present as well.
Such retardation (or non-potential) effects 
start at the next-to-leading order  (NLO) in the multipole expansion and are systematically 
encoded in the theory and  typically related to nonperturbative effects
\cite{Brambilla:1999xf}.
There is a systematic 
procedure to calcolate corrections in $v$ to physical observables:
higher order perturbative (bound state) 
calculations in this framework become viable.
In particular the EFT can be used for a very efficient 
resummation of 
 large logs (typically logs of the ratio of energy and momentum scales) 
 using the renormalization group  (RG) adapted to the case of correlated scales \cite{rg,vnrqcd};
 Poincar\'e invariance is not lost, but shows up 
in some exact relations among the matching coefficients \cite{poincare0}.
The renormalon subtraction 
may be implemented systematically obtaining a perturbative series better behaved 
and allowing a factorization of the genuine QCD nonperturbative effects.

\subsection{Strongly coupled pNRQCD}
In this case the
matching to pNRQCD is nonperturbative. 
Away from threshold
(precisely when heavy-light meson pair and heavy hybrids 
develop a mass gap of order $\lQ$ with respect to the energy of the $Q\bar{Q}$ pair), 
the quarkonium singlet field $\rm S$ remains as the only low energy dynamical 
degree of freedom in the pNRQCD Lagrangian (if no ultrasoft pions are considered), 
which reads \cite{Brambilla:2000gk,Pineda:2000sz,Brambilla:2004jw}:
\begin{equation}
\quad  {\cal L}_{\rm pNRQCD}=  {\rm S}^\dagger
   \left(i\partial_0-{{\bf p}^2 \over
 2m}-V_S(r)\right){\rm S}  .
\end{equation}
The matching potential $V_S(r)$ is a series in the expansion in the inverse of the quark masses:
static, $1/m$ and $1/m^2$ terms have been calculated, see  \cite{Brambilla:2000gk,Pineda:2000sz}.
They involve NRQCD matching coefficients and low energy nonperturbative parts given in terms
of Wilson loops and field strengths insertions in the Wilson loop.
In this regime   we recover the quark potential singlet model from
 pNRQCD.  However the potentials are  calculated from QCD in the formal nonperturbative 
matching procedure. An actual evaluation of the low energy part requires lattice 
evaluation \cite{Bali:2000gf} or QCD vacuum models calculations 
\cite{Brambilla:1999ja,Brambilla:1998bt}.

\section{Applications}

The condition $m\ll \lQ$ always holds and thus the first matching from QCD to 
NRQCD is a perturbative matching. NRQCD describes in principle all heavy quarkonia states
and physical processes. However, since still the soft and the ultrasoft scales  are dynamical,
the power counting is not unambiguous and in some cases may differ from  the perturbative 
inspired BBL counting \cite{Bodwin:1994jh,Brambilla:2002nu,Fleming:2000ib}.
 The number of nonperturbative operators 
tend to increase with the order of the expansion in $v$, and their expectation values
 depend both on the quarkonium states and the US gluons. The NRQCD lattice implementation 
still requires the calculation of the NRQCD matching coefficients in the lattice regularization,
which is still missing in many cases. Being NRQCD a nonrenormalizable theory at the leading order 
Lagrangian, NRQCD lattice calculations maybe tricky. 

The lowest energy EFT, pNRQCD,  is simpler and as such may be more predictive. However, in the present 
formulation, it is valid only for states away from threshold.  Since we are now  integrating 
out also the soft scale, it is important to establish when $\lQ$ sets in, i.e. when we have to 
resort to non-perturbative methods. 
For low-lying resonances, it is reasonable, although not proved, to assume 
$mv^2 \simg \lQ$. The system is weakly coupled and we may rely on perturbation theory,
for instance, to calculate the potential. The theoretical challenge here is performing higher-order 
calculations and the goal is precision physics. For high-lying resonances, we assume
$mv \sim \lQ$.  The system is strongly coupled and the potential must be determined 
non-perturbatively, for instance, on the lattice.  
The theoretical challenge here is providing a consistent framework 
where to perform lattice calculations and the progress is measured by the advance in
 lattice computations. The number of nonperturbative operators maybe be greatly reduced 
with respect to NRQCD, since a further factorization at the soft scale is realized and nonperturbative
 contributions become  typically only a function of the US gluons.
The pNRQCD leading order strongly coupled Lagrangian is renormalizable allowing in principle 
a straightforward lattice implementation.

Both in NRQCD and pNRQCD a source of concern may arise from the large $v^2$ corrections 
in the charmonium case. Large (renormalon-like) perturbative contributions in the matching 
coefficients need to be properly taken care, resummed and subtracted.

\section{QCD potentials}

The QCD potentials achieve a well defined status and definition only in pNRQCD:
they are the matching coefficients of the EFT and as such there is 
a well defined procedure to calculate them.
They depend on the scale of the matching.
In weakly coupled pNRQCD the soft scale is bigger than $\lQ$ and so 
the singlet and octet potentials  have  to be calculated in the 
perturbative matching. In \cite{Brambilla:1999qa}
a determination of the singlet potential at three loops leading log has been obtained 
inside the EFT which gives the way to deal with the well known infrared singularity 
arising in the potential at this order \cite{Appelquist:1977es}.
 From this, $\als$ in the $V$ regularization can be 
obtained, showing a dependence on the infrared behaviour of the theory
at this order and  for this regularization. 
The finite terms in the singlet static potential at three loops are not yet known
but has been estimated \cite{Chishtie:2001mf}. Recently, also the logarithmic 
contribution at four loops has been calculated \cite{Brambilla:2006wp}.
 The three loop renormalization group improved 
calculation of the static singlet potential has been compared to the lattice calculation and found 
in good agreement up to about 0.25 fm \cite{Pineda:2002se}. 
The static octet potential is known at two loops  \cite{Kniehl:2004rk} and again agrees 
well with the lattice data \cite{Bali:2003jq}.

At a scale $\mu$ such that $mv \sim \lQ \gg \mu \gg mv^2$, confinement 
sets in and the potentials become admixture of perturbative 
terms, inherited from NRQCD, which encode high-energy contributions, 
and non-perturbative objects.
Strongly coupled pNRQCD gives us the general form of the  potentials 
obtained in the nonperturbative matching to QCD in the form of Wilson 
loops and Wilson loop chromoelectric and chromomagnetic field strengths insertions
 \cite{Brambilla:2000gk,Pineda:2000sz}, very well suited for lattice calculations.
These will be in general complex valued functions.  The real part controls the spectrum 
and the imaginary part controls the decays.

The real part of the potential has been one of the first quantities to be calculated 
on the lattice (for a review see \cite{Bali:2000gf}).
In the last year, there has been some remarkable progress.
In \cite{Koma:2006si}, the $1/m$ potential has been calculated for the first time.
The existence of this potential was first pointed 
out in the pNRQCD framework \cite{Brambilla:2000gk}. 
A $1/m$  potential is typically missing in potential 
model calculations. The lattice result shows that the potential has a  
$1/r$ behaviour, which, in the charmonium case, 
is of the same size as the $1/r$ Coulomb tail of the static potential 
and, in the bottomonium one, is about 25\%. 
Therefore, if the $1/m$ potential has to be considered part 
of the leading-order quarkonium potential together with the static one, 
as the pNRQCD power counting suggests and the lattice seems to show, 
then  the leading-order quarkonium potential would be, somewhat surprisingly, 
a flavor-dependent function.
In \cite{Koma:2006fw}, spin-dependent potentials have been calculated 
with unprecedented precision. In the long range, they show, for the first 
time, deviations from the flux-tube picture of chromoelectric 
confinement \cite{Buchmuller:1981fr,Brambilla:1998bt}.
The knowledge of the potentials in pNRQCD could provide an alternative to the
direct determination of the spectrum in NRQCD lattice simulations:
the quarkonium masses would be determined by solving the Schr\"odinger
equation with the lattice potentials. The approach may present some advantages: 
the leading-order pNRQCD Lagrangian, differently from the NRQCD one, 
is renormalizable, the potentials are determined once for ever for
all quarkonia, and the solution of the Schr\"odinger equation provides also the 
quarkonium wave functions, which enter in many quarkonium observables: ~
decay widths,~ transitions,~ production cross-sections.
The  existence of a   power counting inside the EFT  selects the 
leading and the subleading terms 
in quantum-mechanical 
perturbation theory. Moreover,  the quantum mechanical  divergences 
(typically encountered in perturbative calculations involving iterations 
of the potentials, as in the case of the  iterations
    of spin delta potentials)  are absorbed  by NRQCD matching coefficients.
Since a  factorization between the   hard (in the NRQCD matching coefficients) and  soft  scales
 (in the Wilson loops or nonlocal gluon correlators) is realized and since 
the  low energy objects are only  glue dependent,
confinement investigations,
        on the lattice 
and in QCD vacuum models become feasible \cite{Brambilla:1999ja,Brambilla:1998bt}.

The potentials evaluated on the lattice once used in the 
Schr\"odinger equation produce the spectrum.
The calculations involve only   QCD parameters  (at some scale and in some 
scheme).

\section{Precision determination of Standard Model parameters}

Given the advancement in the EFTs formulation and in the lattice calculations
 as well as  the existence of several high order perturbative bound state calculations, 
quarkonia may become  a very appropriate system for the extraction of precise determination 
of the Standard Model parameters like $\als$ and the heavy quark masses. Such precise determinations 
are important for physics inside and beyond the Standard Model.
Inside the QWG (www.qwg.to.infn.it)
 there is a topical group for  such studies and in the QWG YR \cite{Brambilla:2004wf}
there is a dedicated chapter.

\subsection{$c$ and $b$ mass extraction}
The lowest  heavy quarkonium states are   suitable systems to extract a precise 
determination of the mass of the heavy quarks $b$ and $c$.
Perturbative determinations of the 
$\Upsilon(1S)$ and $J/\psi$ masses have been used to extract the $b$ and $c$
masses. The main uncertainty in these determinations
comes from nonperturbative nonpotential contributions (local and nonlocal condensates)
together with possible effects due to
subleading renormalons. 
These determinations  are competitive with those coming from different systems 
and different approaches (for the $b$ mass see e.g. \cite{El-Khadra:2002wp}).
We report some recent determinations in Tab. \ref{Tabmasses}.

\begin{table}[h]
\addtolength{\arraycolsep}{0.2cm}
\begin{center}
\begin{tabular}{|c|c|c|}
\hline
reference & order &  ${\overline m}_b({\overline m}_b)$ (GeV) 
\\
\hline
\cite{Pineda:2001zq} & NNNLO$^*$ & $4.210 \pm 0.090 \pm 0.025$
\\
\cite{Brambilla:2001qk} & NNLO +charm & $4.190 \pm 0.020 \pm 0.025$
\\
\cite{Eidemuller:2002wk} & NNLO& $4.24 \pm 0.10$
\\
\cite{Penin:2002zv} & NNNLO$^*$& $4.346 \pm 0.070$
\\
\cite{Lee:2003hh} & NNNLO$^*$ & $4.20 \pm 0.04$
\\
\cite{Contreras:2003zb} & NNNLO$^*$ & $4.241 \pm 0.070$ 
\\
\cite{Pineda:2006gx} & NNLL$^*$ & $4.19 \pm 0.06$ 
\\
\hline
\hline
reference  & order &  ${\overline m}_c({\overline m}_c)$ (GeV)  
\\
\hline
\cite{Brambilla:2001fw} & NNLO & $1.24 \pm 0.020$ 
\\
\cite{Eidemuller:2002wk}  & NNLO & $1.19 \pm 0.11$ 
\\
\hline
\end{tabular}
\vspace{2mm}
\caption{Different recent determinations of ${\overline m}_b({\overline m}_b)$
and ${\overline m}_c({\overline m}_c)$ in the $\MS$ scheme from the bottomonium and the 
charmonium systems. The displayed results either use direct determinations or non-relativistic 
sum rules. Here and in the text, the $^*$ indicates that the theoretical input is only partially 
complete at that order.}
\end{center}  
\label{Tabmasses}
\end{table} 

A recent analysis  performed by the QWG \cite{Brambilla:2004wf}
and based on all the previous determinations indicates
 that at the moment the mass extraction from  heavy quarkonium 
involves  an error of about 50 MeV both for the  bottom 
($1\% $ error) and in the charm ($4 \% $ error) mass. It would be 
very important to be able to further reduce the error on the heavy quark masses.

\subsection{Determinations of  $\als$.}
Heavy quarkonia leptonic and non-leptonic inclusive and radiative decays 
 may provide means to extract $\alpha_s$. The present 
  PDG determination of $\alpha_s$ from bottomonium pulls down 
  the global $\alpha_s$ average  noticeably \cite{Brambilla:2004wf}. 
  Recently, using the most recent CLEO data on radiative $\Upsilon(1S)$ decays 
   and dealing with the octet contributions within weakly coupled pNRQCD,
a new determination of $\als(M_{\Upsilon(1S)})= 0.184^{+0.014}_{-0.013}$
has been obtained \cite{Brambilla:2007cz}, which corresponds to 
$\als(M_Z)= 0.119^{+0.006}_{-0.005}$ in agreement 
  with the central value of the PDG \cite{Yao:2006px}
 and with competitive errors.

\subsection{Top-antitop production  near threshold at ILC.}
In \cite{Pineda:2006gx,Hoang:2001mm}
the total cross section for top quark pair production 
close to threshold in e+e- annihilation is investigated at NNLL in the weakly coupled
EFT. The summation of the large logarithms in the ratio of the energy scales
 is achieved with the renormalization group (for correlated scales) and 
significantly reduces the scale dependence of the results.
Studies  like these will make  feasible a precise extractions 
of the strong coupling, the top mass 
and the top width at a future ILC.

\section{Spectra}

The NRQCD Lagrangian is well suited for lattice calculations \cite{Lepage:1992tx}. 
The quark propagators 
are the nonrelativistic ones and since  
the heavy-quark mass scale has been integrated out, 
for NRQCD on the lattice, it is sufficient to have a 
lattice spacing $a$ as coarse as $m \gg 1/a \gg mv$.
A price to pay is that, by construction, the continuum limit cannot be reached. 
Another one is that the NRQCD Lagrangian has to be supplemented by matching coefficients 
calculated in lattice perturbation theory, which encode the contributions from the 
heavy-mass energy modes that have been integrated out. A recent unquenched determination 
of the bottomonium spectrum with staggered sea quarks can be found in \cite{Gray:2005ur}.
The fact that all matching coefficients of NRQCD on the lattice are taken at their tree-level value 
induces a systematic error of order $\als v^2$ for the radial splittings 
and of order $\als$ for the fine and hyperfine splittings. 

Inside pNRQCD we have to  consider separately systems with a small interquark radius (low-lying states) 
and systems with a radius comparable or bigger than the confinement scale 
$\lQ^{-1}$ (high-lying states). It is difficult to say to which group a specific 
resonance may belong, since there are no direct measurements of the interquark radius.
Electric dipole transitions or quarkonium dissociation in a medium, once a 
well founded theory treatment of such processes will be given, may give a clear cut procedure.
 At the moment one uses the typical EFT approaches assuming  that a particular scales hierarchy 
holds  and checking then a posteriori that the prediction and the error estimated inside 
such framework are consistent with the data.

Low-lying $Q\bar{Q}$ states are assumed to realize the hierarchy: $ m \gg  mv
\gg mv^2 \simg \lQ$ and they may be described in weakly coupled pNRQCD.

\begin{table}[ht]
\begin{tabular}{|c|cccc|}
\hline
\multicolumn{5}{|c|}{$B_c$ mass ~(MeV)}\\
\hline
\cite{Acosta:2005us} (expt) &\cite{Allison:2004be} (lattice) & \cite{Brambilla:2000db} (NNLO)
& \cite{Brambilla:2001fw} (NNLO)& \cite{Brambilla:2001qk} (NNLO)\\
\hline
$6287\pm 4.8 \pm 1.1 $ &$6304\pm12^{+12}_{-0}$ & 6326(29) & 6324(22) & 6307(17) 
\\
\hline
\end{tabular}
\caption{Different perturbative determinations of the $B_c$ mass compared with the experimental 
value and a recent lattice determination.}
\label{TabBc}
\end{table}

Once the heavy quark masses are known, one may use them to extract other
quarkonium ground-state observables. The $B_c$ mass has been  calculated 
at NNLO  in \cite{Brambilla:2000db,Brambilla:2001fw,Brambilla:2001qk},
see Table \ref{TabBc}. 
These values 
agree well with the unquenched NRQCD/Fermilab method) lattice determination of \cite{Allison:2004be}, 
which shows that the $B_c$ mass
is not very sensitive to non-perturbative effects. 
This is confirmed by a recent measurement of the $B_c$ in the channel  $B_c \to J/\psi \, \pi$ 
by the CDF collaboration at the Tevatron; they obtain with 360 pb$^{-1}$ of data 
$M_{B_c} = 6285.7 \pm 5.3 \pm 1.2$ MeV \cite{Acosta:2005us}, while the latest
available figure based on 1.1 fb$^{-1}$ of data is 
$M_{B_c} = 6276.5 \pm 4.0 \pm 2.7$ MeV 
(see http://www-cdf.fnal.gov/physics/new/bottom/060525.blessed-bc-mass/).

The bottomonium and charmonium ground-state hyperfine splitting has been calculated at NLL
in \cite{Kniehl:2003ap}. Combining it with the measured $\Upsilon(1S)$ mass,
this determination provides a quite precise prediction for the $\eta_b$ mass: 
$M_{\eta_b} = 9421 \pm 10^{+9}_{-8} ~{\rm MeV}$, where the first error is an estimate of the
theoretical uncertainty and the second one reflects the uncertainty in $\als$.
Note that the discovery of the $\eta_b$ may provide a very competitive 
source of $\als$ at the bottom mass scale with a projected error at the $M_Z$ 
scale of about $0.003$. Similarly, in \cite{Penin:2004xi}, 
the hyperfine splitting of the $B_c$ was calculated at NLL
accuracy: $M_{B_c^*}  - M_{B_c} = 65 \pm 24^{+19}_{-16}~{\rm MeV}$.

High-lying $Q\bar{Q}$ states are assumed to realize the hierarchy: $ m \gg  mv \sim \lQ
\gg mv^2$. A first question is where the transition from low-lying to high-lying takes place.
This is not obvious, because we cannot measure directly $mv$. Therefore, the 
answer can only be indirect and, so far, there is no clear agreement in the literature.
A weak-coupling treatment for the lowest-lying 
bottomonium states ($n=1$, $n=2$ and also for the $\Upsilon(3S)$) appears 
to give positive results for the masses  at NNLO in \cite{Brambilla:2001fw}
and at N$^3$LO$^*$ in \cite{Penin:2005eu}.
The result is more ambiguous for the fine splittings of the 
bottomonium $1P$ levels in the NLO analysis of \cite{Brambilla:2004wu} and positive only 
for the $\Upsilon(1S)$ state in the  N$^3$LO$^*$ analysis of \cite{Beneke:2005hg}. 
 
Masses of high-lying quarkonia may be accessed either using the lattice nonperturbative 
potentials inside a Schr\"odinger equation \cite{Bali:1997am}
 or via a direct lattice pNRQCD calculation.

\section{Transitions and decays}

\subsection{Inclusive Decays}

NRQCD gives a factorization formula for heavy
quarkonium  inclusive decay widths,
precisely it factorizes four-fermion matching coefficients and  matrix 
elements of four fermion operators \cite{Bodwin:1994jh}.
Color singlet operator expectation values may be easily related to the square 
of the quarkonium wave functions (or derivatives of it) at the origin. 
Octet contributions remain as  nonperturbative matrix elements 
of operators evaluated over the quarkonium states.
In some situations the octet contributions may not be suppressed and become as 
  relevant as the  singlet contributions in the NRQCD power counting. In particular  
octet contributions 
may reabsorb the dependence on the infrared cut-off  of the Wilson coefficients, 
solving the problem that arised originally in the color singlet potential model
\cite{Barbieri:1980yp}.

Systematic improvements are possible, either by calculating higher-order corrections 
in the coupling constant or by adding higher-order operators.

In order to describe electromagnetic and hadronic inclusive decay widths 
of heavy quarkonia, many NRQCD matrix elements are needed. 
The specific number depends on the order in $v$ of the non-relativistic 
expansion to which the calculation is performed and on the power counting.
At order $mv^5$ and within a conservative power counting, 
$S$- and $P$-wave electromagnetic and hadronic decay widths 
for bottomonia and charmonia below threshold 
depend on 46 matrix elements \cite{Brambilla:2002nu}.
More are needed at order $mv^7$ \cite{Bodwin:2002hg,Ma:2002ev,Brambilla:2006ph}.
Order $mv^7$ corrections are particularly relevant for $P$-wave quarkonium 
decays, since they are numerically as large as NLO corrections 
in $\als$, which are known since long time \cite{Barbieri:1980yp} and to which the most recent 
data are sensitive \cite{Vairo:2004sr,Brambilla:2004wf}.
NRQCD matrix elements may be fitted to the experimental decay data
\cite{Mangano:1995pu,Maltoni:2000km} or calculated on the lattice
\cite{Bodwin:1996tg,Bodwin:2001mk,Bodwin:2005gg}. The matrix elements of
color-singlet operators are related at leading order to the
Schr\"odinger wave functions at the origin \cite{Bodwin:1994jh} 
and, hence, may be evaluated by means of potential models
\cite{Eichten:1995ch} or potentials calculated on the lattice
\cite{Bali:2000gf}.  However, a great  part of the matrix elements remain 
poorly known or unknown.

In the matching coefficients large contributions in the perturbative series 
coming from bubble-chain diagrams may need to be resummed \cite{chen}.

In lattice NRQCD in \cite{Gray:2005ur}, 
the ratio 
$\Gamma(\Upsilon(2S)\to e^+e^-)/\Gamma(\Upsilon(1S)\to e^+e^-)\times M_{\Upsilon(2S)}^2/M_{\Upsilon(1S)}^2$ 
has been calculated. The result on the finest lattice compares well with the 
experimental one.

The imaginary part of the potential provides the NR\-QCD decay matrix elements
in pNRQCD. 
For excited states, they typically factorize in a part, which is the wave function in
the origin squared (or its derivatives), and in a part which contains 
gluon tensor-field correlators 
\cite{Brambilla:2001xy,Brambilla:2002nu,Brambilla:2003mu,Vairo:2003gh}.
This drastically reduces the number of non-perturbative parameters needed; in pNRQCD,
these are wave functions at the origin and universal gluon tensor-field correlators, 
which can be calculated on the lattice.
Another approach may consist in determining the correlators on one set 
of data (e.g. in the charmonium sector) and use them to make predictions 
for another (e.g. in the bottomonium sector). Following this line in 
\cite{Brambilla:2001xy,Vairo:2002nh}, at NLO in $\als$, but at leading 
order in the velocity expansion, it was predicted 
${\Gamma_{\rm had}(\chi_{b0}(2P))}/{\Gamma_{\rm had}(\chi_{b2}(2P))} \approx$
 $4.0$ and ${\Gamma_{\rm had}(\chi_{b1}(2P))}/$ ${\Gamma_{\rm
had}(\chi_{b2}(2P))} \approx$ $0.50$. Both determinations turned 
out to be consistent, within large errors, with the CLEO III data 
\cite{Brambilla:2004wf}. One should notice that at some order of the expansion 
in $v$,  the scale $\sqrt{m \lQ}$ start also to contribute in pNRQCD jeopardizing in some cases
the effective reduction of the nonperturbative operators \cite{Brambilla:2003mu}.

For the lowest resonances, inclusive decay widths are given in weakly coupled pNRQCD by a 
convolution of perturbative corrections and nonlocal nonperturbative correlators.
The perturbative calculation embodies large contributions and 
requires large logs resummation.
The ratio of electromagnetic decay widths was calculated for the ground state 
of charmonium and bottomonium at NNLL order in \cite{Penin:2004ay}.
In particular, they report: $\Gamma(\eta_b\to\gamma\gamma) /\Gamma(\Upsilon(1S)\to e^+e^-) = 0.502
\pm 0.068 \pm 0.014$, which is a very stable result with respect to scale variation.
A partial  NNLL$^*$ order analysis of the absolute width of $\Upsilon(1S) \to
e^+e^-$ can be found in \cite{Pineda:2006ri}.

\subsection{Electromagnetic transitions}

Allowed magnetic dipole transitions between charmonium and bottomonium ground states
have been considered in pNRQCD at NNLO  in \cite{Brambilla:2005zw,Vairo:2006js}.
The results are: $\Gamma(J/\psi \to \gamma \, \eta_c) \! = (1.5 \pm 1.0)~\hbox{keV}$
and $\Gamma(\Upsilon(1S) \to \gamma\,\eta_b)$ $=$  $(k_\gamma/39$ $\hbox{MeV})^3$
$\,(2.50 \pm 0.25)$ $\hbox{eV}$, where  the errors account for uncertainties (which
are large in the charmonium case) coming from higher-order corrections.
The width $\Gamma(J/\psi \to \gamma\,\eta_c)$ is consistent with 
\cite{Yao:2006px}. Concerning $\Gamma(\Upsilon(1S) \to \gamma\,\eta_b)$, 
a photon energy $k_\gamma = 39$ MeV corresponds to a $\eta_b$ mass of 9421 MeV. 
The pNRQCD calculation features  a small quarkonium magnetic moment (in agreement 
with a recent lattice calculation 
\cite{Dudek:2006ej}) and the interesting 
fact, related to the Poincar\'e invariance of the nonrelativistic 
EFT \cite{Brambilla:2003nt}, 
that M1 transition of the lowest quarkonium states at relative order $v^2$ are 
completely accessible in perturbation theory \cite{Brambilla:2005zw}.

In the weak-coupling regime, the magnetic-dipole hindered transition 
$\Upsilon(2S) \to \gamma\,\eta_b$ 
at leading order \cite{Brambilla:2005zw} does not agree with the experimental upper bound 
\cite{Artuso:2004fp}. It should be still clarified if this is related to the fact 
that $\Upsilon(2S)$ system belongs to the strong coupling regime or if it is due to 
large corrections (more relevant in the hindered case).

\subsection{Semi-inclusive decays}

The radiative transition $\Upsilon(1S)\to\gamma\,X$ has been considered in 
\cite{Fleming:2002sr,GarciaiTormo:2005ch}. The agreement with the CLEO data of 
\cite{Nemati:1996xy} is very satisfactory when one properly includes the octet 
contribution in pNRQCD \cite{GarciaiTormo:2005ch}.
 In the same work it is found that the ratios for different $n$ of the radiative decay widths 
$\Gamma(\Upsilon(nS) \to \gamma\,X)$ are better consistent with
the data if $\Upsilon(1S)$ is assumed to be a weakly-coupled bound state 
and $\Upsilon(2S)$ and $\Upsilon(3S)$ strongly coupled ones \cite{GarciaiTormo:2005bs}.

 In general in exclusive decays 
and for certain kinematical end points of semi-inclusive decays, NRQCD or pNRQCD 
should be supplemented by collinear degrees of freedom. This can be realized 
in the framework of Soft Collinear Effective Theory (SCET) \cite{Bauer:2000yr}.

\section{Baryons with two or more heavy quarks}

The SELEX collaboration at Fermilab reported evidence for  five resonances that 
may be possibly  identified with doubly charmed baryons, see the presentation 
of Jurgen Engelfried at this meeting \cite{here} and \cite{Ocherashvili:2004hi}. 
Although these results have not been confirmed by other experiments ( 
 FOCUS, BELLE and BABAR) they have triggered a renewed 
theoretical interest in doubly heavy baryon systems.

Low-lying $QQq$ states may be  assumed to realize the hierarchy: $ m \gg mv \gg
\lQ$, where $mv$ is the typical inverse distance between the two heavy quarks
and $\lQ$ is the typical inverse distance between the centre-of-mass of the two heavy quarks
and the light quark. 
At a scale $\mu$ such that $mv \gg \mu \gg \lQ$ the effective  
degrees of freedom are $QQ$ states (in color antitriplet and sextet
configurations), low-energy gluons and light quarks. The most suitable EFT at
that scale is a combination of pNRQCD and HQET
\cite{Brambilla:2005yk,Fleming:2005pd}. The hyperfine splittings of the doubly heavy 
baryon lowest states have been calculated at NLO in $\als$ and at LO in
$\lQ/m$ by relating them to the hyperfine splittings of the $D$ and $B$ mesons (this 
method was first proposed in \cite{Savage:di}). In \cite{Brambilla:2005yk}, the 
obtained values are: $M_{\Xi^*_{cc}}-M_{\Xi_{cc}} = 120 \pm 40$ MeV 
and $M_{\Xi^*_{bb}}-M_{\Xi_{bb}} = 34 \pm 4$ MeV, which are 
consistent with the quenched lattice determinations of 
\cite{Flynn:2003vz,Lewis:2001iz} and the quenched NRQCD lattice 
determinations of \cite{AliKhan:1999yb,Mathur:2002ce}.
Chiral corrections to the doubly  heavy baryon masses, strong decay widths and 
electromagnetic decay widths have been considered in \cite{Hu:2005gf}.

Also low-lying $QQQ$ baryons can be studied in a weak coupling framework.
Three quark states can combine in four color configurations: a singlet,
two octets and a decuplet, which lead to a rather rich dynamics
\cite{Brambilla:2005yk}. Masses of various $QQQ$ ground states have been 
calculated with a variational method in \cite{Jia:2006gw}: since baryons made of three 
heavy quarks have not been discovered so far, it may be important for future searches  
to remark that the baryon masses turn our to be lower 
than those generally obtained in strong coupling analyses.

For $QQQ$ baryons with a typical distance of the order $\lQ$ inverse, the form of the 
static, $1/m$ and spin dependent nonperturbative potentials have been obtained in pNRQCD
\cite{Brambilla:2005yk}. Up to now only the static potential has been evaluated on the lattice
\cite{Bali:2000gf,Suganuma:2004zx}.

\section{Gluelump spectrum and exotic states}
Gluelumps are states formed by a gluon and two heavy quarks in a octet configuration 
at small interquark distance \cite{Foster:1998wu}  . 
The mass of such nonperturbative objects are typically measured on the lattice. The 
tower of hybrids static energies \cite{Juge:2002br} measured in lattice NRQCD
 reduces to the gluelump masses for small interquark distances.
In pNRQCD \cite{Brambilla:1999xf,Bali:2003jq}  
 the full structure of the gluelump spectrum has been studied,
obtaining model independent predictions on the shape, the pattern, the degeneracy 
and the multiplet structure of the hybrid static energies for small $Q\bar{Q}$ 
distances that well match and interpret the existing lattice data.  
These studies may be important both to  elucidate the confinement mechanism (the gluelump 
masses control the behaviour of the nonperturbative glue correlators appearing in the spectrum 
and in the decays) and in relation to the exotic states recently observed at the 
B-factories. The $Y(4260)$  in the charmonium sector may be 
identified with an hybrid state inside such picture.
A complete pNRQCD description of heavy hybrids is still missing.

\section{Production}
Before the advent of NRQCD, colour singlet production and colour singlet 
fragmentation underestimated the data on prompt quarkonium production at Fermilab 
by about an order of magnitude indicating that additional fragmentation contributions 
were missing \cite{Kramer:2001hh}.
The missing contribution was precisely the gluon fragmentation into 
colour-octet $ ^3S_1$ charm quark pairs. The probability to form a $J/\psi$ particle from a pointlike 
$c\bar{c}$  pair in a colour octet  $ ^3S_1$ state is given by a NRQCD 
 nonperturbative matrix element 
which is suppressed by $v^4$ with respect to  to the leading singlet term but 
 is enhanced by two powers of $\als$ 
 in the short distance matching coefficient  
for producing colour-octet quark pairs. Introducing  the leading 
colour-octet contributions, the data of CDF could  be reproduced and this has 
been an important result of NRQCD \cite{Kramer:2001hh}.
 NRQCD factorization has proved to be very successful 
to explain a large variety of quarkonium production processes (for a review see 
the production chapter in \cite{Brambilla:2004wf} and 
\cite{nreftrev2}).
A  formal proof of the NRQCD factorization formula for heavy quarkonium production 
 has however not yet been obtained. Recently, there has been  important work 
in the direction of an all order proof
in \cite{Nayak:2005rt,Nayak:2006fm}. In particular,  it has been 
shown that a necessary condition 
for factorization to hold at NNLO is that the conventional octet NRQCD production matrix elements 
have to be redefined with Wilson lines, acquiring  manifest gauge invariance.

For production, a pNRQCD formulation is not yet existing. In principle to go 
through a further factorization also in production, if at all possible, 
may reduce the number of nonperturbative matrix elements and enhance the predictive power.

Two outstanding problems exist at the moment in quarkonium production:
 double charmonium production in $e^+e^-$ collisions and 
 charmonium polarization at the Tevatron.

In \cite{Abe:2004ww}, BELLE reports
$\sigma(e^+e^-\to J/\psi+\eta_c) \, {\rm Br}(c\bar{c}\to$ $> \hbox{2 charged}) 
= 25.6\pm 2.8\pm 3.4~\hbox{fb}$ and in \cite{Aubert:2005tj}, BABAR reports 
$\sigma(e^+e^-\to J/\psi+\eta_c) \, {\rm Br}(c\bar{c} \to$ $> \hbox{2 charged}) 
= 17.6\pm 2.8^{+1.5}_{-2.1}~\hbox{fb}$. Originally these data 
were about one order of magnitude above theoretical expectations. 
Recently, with some errors corrected in some of the theoretical 
determinations,   NLO corrections in $\als$ calculated in 
\cite{Zhang:2005ch} and  higher-order $v^2$ corrections obtained in \cite{Bodwin:2006dn},
the theoretical prediction has moved   closer to 
the experimental one. In \cite{Lee:QWG2006}, a preliminary 
estimate of $\sigma(e^+e^-\to J/\psi+\eta_c)$ including the above corrections 
gives $ 16.7\pm 4.2~\hbox{fb}$.

Still open is the problem of the BELLE measurement
$\sigma(e^+e^-\to J/\psi+ c\bar{c})/\sigma(e^+e^-\to J/\psi+ X)$,  about 
80\%, with respect to theory  calculation  that gives about 10\% 
(see \cite{Brambilla:2004wf} for a detailed discussion). 

Charmonium polarization has been measured at the Tevatron by the CDF collaboration 
at run I with 110 pb$^{-1}$  \cite{Affolder:2000nn} and recently at run II 
with  800 pb$^{-1}$  \cite{Kim:QWG2006}.
The data of the two runs are not consistent with each other 
in the 7-12 GeV region of transverse momentum, $p_T$, and both 
disagree with the  NRQCD expectation of an increased polarization with 
increased $p_T$.
For large $p_T$, NRQCD predicts that 
the main mechanism of charmonium production is via color-octet gluon fragmentation, 
the gluon being transversely polarized and most of the gluon polarization 
being transferred to the charmonium.
The CDF data do not show any sign of transverse polarization at large $p_T$.

A solution to such problem may be obtained in the case in which a 
nonperturbative power counting is valid
\cite{Fleming:2000ib,Brambilla:2002nu}.

\section{Challenges}

For what concerns systems close or above the open flavor threshold, 
a complete and satisfactory understanding of the dynamics has not been achieved so far
and a corresponding general NR EFT has not yet been constructed. Such systems 
are difficult to address also with a lattice calculation. 
Hence, the study of these systems is on a less secure 
ground than the study of states below threshold.  
Although in some cases one may develop an EFT owing to special dynamical conditions
(as for the $X(3872)$ interpreted as a 
loosely bound $D^0 \, \bar{D}^{*\,0}$ $+$ ${\bar D}^0 \, D^{*\,0}$ molecule
\cite{Braaten:2003he}), 
the study of these systems largely relies on phenomenological models 
\cite{Eichten:2005ga,Swanson:2006st}.
The major theoretical challenge here is to interpret the new 
states in the charmonium region discovered at the B-factories in the last few years. 

Heavy ion experiments use quarkonium suppression as one of the smoking guns 
for quark-gluon plasma formation (cf. e.g. the media chapter in \cite{Brambilla:2004wf}).
. To describe quarkonium suppression it would be 
important to formulate an EFT for heavy quarkonium in media and to obtain a 
clear definition of the heavy quark potential at finite T . Preliminary 
studies are ongoing with several approaches \cite{Jakovac:2006sf}.

With a good control in theory and high statistic data sample 
available at present and future (Super-B factory) experiments,
heavy quarkonia may also supply us with an alternative way  of 
looking for new physics BSM, cf. \cite{Sanchis-Lozano:2006gx} and the 
BSM chapter in \cite{Brambilla:2004wf}.

\section{Outlook}

Today NR EFTs and lattice calculations allow us
to investigate a wide range of heavy quarkonium observables 
in a controlled and systematic fashion and, therefore, learn about one of the most elusive sectors 
of the Standard Model: low-energy QCD.
 
Predictions based on non-relativistic EFTs 
are conceptually solid, and systematically improvable. 
EFTs have put quarkonium on the solid ground of QCD: 
quarkonium becomes a  privileged window  for precision measurements,
 new physics and  confinement mechanism investigations.

Many new data on heavy-quark bound states are provided in these years by the 
B-factories, CLEO and the Tevatron experiments. Many more will come 
in the near future from BES-III,  LHC and later Panda at GSI. 
They will show new (perhaps exotic) states, new production and decay mechanisms and they 
will be a great arena for new  EFT tools.

\section{Acknowledgments}
I would like to thank the organizers for the invitation and for the perfect 
organization of this very enjoable conference. 
The support by the European Research Training 
Network Flavianet, contract number MRTN-CT-2006-035482 is gratefully acknowleged.

\end{document}